\documentclass[twocolumn,prb]{revtex4}
\usepackage{graphicx}
\usepackage{bm}
\usepackage{amsmath}
\begin{document}

\title{The Silicon Inversion Layer With A Ferromagnetic Gate: A Novel
Spin Source}

\author{J. P. McGuire, C. Ciuti, L. J. Sham}

\affiliation{Department of Physics, University of California, San
  Diego, La Jolla, CA 92093-0319.}

\begin{abstract}
Novel spin transport behavior is theoretically shown to result from
replacing the usual metal (or poly-silicon) gate in a silicon
field-effect transistor with a ferromagnet, separated from the
semiconductor by an ultra-thin oxide.  The spin-dependent interplay
between the drift current (due to a source-drain bias) and the
diffusion current (due to carrier leakage into the ferromagnetic gate)
results in a rich variety of spin dependence in the current that flows
through such a device.  We examine two cases of particular interest:
(1) creating a $100\%$ spin-polarized electrical current and (2)
creating a pure spin current without a net electrical current.  A
spin-valve consisting of two sequential ferromagnetic gates is shown
to exhibit magnetoresistance dependent upon the relative orientations
of the magnetization of the two ferromagnets.  The magnetoresistance
ratio grows to arbitrarily large values in the regime of low
source-drain bias, and is limited only by the spin-flip time 
in the channel.
\end{abstract}


\date{\today}

\maketitle

The creation, manipulation, and detection of spin-polarized carriers
are the essential ingredients of semiconductor
spintronics.\cite{review1, review2, review3} Although well established
using optical methods,\cite{opt_orientation, kikkawa, gupta} the
ultimate goal is to achieve this level of control over the spin degree
of freedom in all-electrical devices.\cite{datta} To this end, spin
injection from ferromagnets into semiconductors\cite{fiederling, ohno,
zhu, hanbicki} and spin control through the Rashba spin-orbit
effect\cite{rashba, nitta1, nitta2, marcus1, marcus2} have attracted
much interest during the last few years.  In addition, it has recently
been proposed\cite{bhat} and experimentally verified\cite{stevens,
hubner} that a {\it pure} spin current, without an accompanying
electrical current, can be created in direct-gap semiconductors using
the interference of one- and two-photon absorption processes.  Such
pure spin currents will allow for new probes of spin-dependent
phenomena in semiconductors and may lead to alternate spintronics
device designs not present in existing charge-based technology.

In this paper we present a method of creating fully spin-polarized
charge currents and pure spin currents using a ferromagnetically-gated
inversion layer in the regime of low source-drain bias.\cite{apl, prb}
Our method is possible in any semiconductor, notably silicon, and not
just those with well-defined optical transitions.  It relies on a
proximity effect in which semiconductor carriers acquire
spin-dependent properties through coupling with a nearby
ferromagnet.\cite{prl}   Experiments have
shown that optically-pumped unpolarized electrons in a semiconductor
acquire a net spin-polarization in the presence of an interface with a
ferrromagnet.\cite{kawakami, epstein1, epstein2}
The magnitude of
the spin-dependent effects are optimized by confining the
semiconductor carriers at the interface with the ferromagnet, with the
ferromagnet acting as the electrically-biased gate responsible for the
inversion.  The crucial parameter is the thickness of the oxide
barrier which separates the semiconductor from the ferromagnet; since
the coupling of the two depends exponentially on their separation, our
proposal will be most relevant for ultra-thin gate oxides at the
forefront of current technology.\cite{momose}  The balance of growing oxides thin
enough to produce ample coupling while preserving the functionality of
the inversion layer is delicate, although we point out that we are
merely taking advantage of the aggressive scaling of field-effect
transistors to the nanoscale.  In addition, we will discuss the
behavior of a recently-proposed spin-valve with two sequential
ferromagnetic gates in the same regime of low source-drain bias, where
the magnetoresistance ratio due to the relative orientation of the
ferromagnet magnetizations can grow to arbitarily large values.

\begin{figure}[]
\includegraphics[width=8cm]{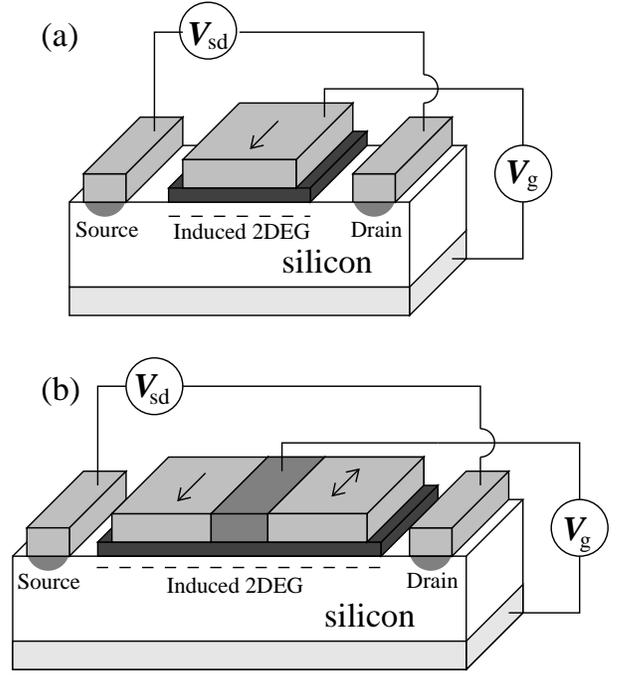}
\caption{ (a) The silicon field-effect transistor with a single
ferromagnetic gate.  The source-drain bias $V_{\text{sd}}$ creates the
in-plane field $E_x$ that drives the drift current.  The ferromagnetic
gate is biased ($V_{\text g}$) to create the inversion layer.  (b) The
spin-valve with two adjacent ferromagnetic gates; the two ferromagnets
have different coercive fields to enable the switching of the second
gate's magnetization without affecting that of the first gate.  The
gap between the two ferromagnets is filled with a paramagnetic metal
to ensure uniform 2DEG confinement throughout the device.}
\label{device_fig}
\end{figure}

To demonstrate the operational principle of the device, consider the
spin-dependent current in a two-dimensional electron gas (2DEG).  We
will concentrate on the silicon inversion layer with an
${\text{SiO}}_2$ barrier and a ferromagnetic gate, and consider only a
single occupied subband in the silicon.  For simplicity, we will
assume the two spin channels are completely decoupled, but we will
return to address any coupling between the two spin channels later in
the paper.  The device is shown in Fig.~\ref{device_fig}(a); the
growth axis is in the $\hat{z}$-direction, the in-plane current flows
in the $\hat{x}$-direction, and we assume the device is
translationally invariant in the $\hat{y}$-direction.  The in-plane
electrical current $J_\sigma(x)$ in spin channel $ \sigma = \{
\uparrow,\downarrow \} $ contains both drift and diffusion terms,
\begin{equation}
J_\sigma(x) = g_\sigma(x) E_x + e D_\sigma(x) \frac{\partial}
{\partial x} N_\sigma(x)~,
\label{current}
\end{equation}
where $-e$ is the electron charge, $g_\sigma(x)$ is the Drude
conductivity, $E_x$ is the electric field that drives the in-plane
current, $D_\sigma(x)$ is the diffusion constant, and $N_\sigma(x)$ is
spatially-dependent 2DEG density.  To achieve spin-dependent
transport, either the transport coefficients or the 2DEG density must
vary considerable between the two spin channels.  We show below how
the coupling between the 2DEG and a ferromagnetic gate achieves this
splitting.

In Ref.~\onlinecite{apl} the coupling was calculated from the
wavefunction matching conditions, resulting in an equation which
specified the spin-dependent complex energy of the confined inversion
layer state coupled to the ferromagnet.  In Ref.~\onlinecite{prb}
an effective tight-binding Hamiltonian was derived from the
effective-mass Hamiltonian of the silicon/${\text{SiO}}_2$/ferromagnet
interface.  The form of this Hamiltonian allowed for the calculation
of the 2DEG self-energy due to the interaction with the ferromagnet,
the approach we will adopt in this paper.  The results of the
derivation in Ref.~\onlinecite{prb} are summarized below in order
to clarify the method of calculation used throughout the rest of this
paper.  The effective Hamiltonian of the 2DEG coupled to the
ferromagnetic gate is
\begin{eqnarray}
\nonumber H &=& \sum_\sigma a_\sigma^\dagger \epsilon^{\text{si}}
a_\sigma + \sum_{k,\sigma} c_{k,\sigma}^\dagger
\epsilon^{\text{fm}}_{k,\sigma} c_{k,\sigma} \\ &~& + \sum_{k,\sigma}
\left [ a_\sigma^\dagger V^{\text{fm}}_{k,\sigma} c_{k,\sigma} +
c_{k,\sigma}^\dagger \left ( V^{\text{fm}}_{k,\sigma} \right )^\star
a_\sigma \right ]~.
\label{Hamiltonian}
\end{eqnarray} 
A one-dimensional calculation is possible for the coupling calculation
because the conservation of the in-plane wavevector is
relaxed.\cite{lo, shih} The operator $a_\sigma^\dagger$ creates the
inversion layer state with spin $\sigma$, energy
$\epsilon^{\text{si}}$, and wavefunction $\chi^{\text{si}}(z)$, which
is the lowest bound state of the potential
\begin{equation}
V^{\text{si}}(z) = \left [ -\tilde{U}_{\text b} - eE^{\text{si}}_z
  \cdot \left ( z - z_{\text b} \right ) \right ] \Theta(z-z_{\text
  b})~,
\end{equation}
where $\tilde{U}_{\text b}$ is the oxide barrier height from the
bottom of the semiconductor potential, $E^{\text{si}}_z$ is the
electric field in the semiconductor, and $z_{\text b}$ is the
thickness of the oxide barrier.  The "0" of energy has been put at the
top right side of the barrier to simplify the calculations.  The
operator $c_{k,\sigma}^\dagger$ creates a ferromagnet state with spin
$\sigma$, wavevector $k$, energy $\epsilon^{\text{fm}}_{k,\sigma}$,
and wavefunction $\phi^{\text{fm}}_{k,\sigma}(z)$, which are states of
the potential
\begin{eqnarray}
\nonumber V^{\text{fm}}_\sigma(z) &=& - \left [
U_{\text{fm}}+\frac{\Delta}{2} {\hat{\bm{\sigma}}} \cdot
{\hat{\bm{M}}} \right ] \Theta(-z) \\ &&- e E^{\text b}_z \cdot
(z-z_{\text b}) \Theta(z) \Theta(z_ {\text b}-z)~,
\end{eqnarray}
where $U_{\text{fm}}+\Delta/2$ is the Fermi energy of the majority
spin band in the ferromagnet, $U_{\text{fm}}-\Delta/2$ is the Fermi
energy of the minority spin band in the ferromagnet, and $E^{\text
b}_z$ is the electric field in the barrier.  Note that for simplicity
we treat the ferromagnet as exchange-split parabolic bands.\cite{slon}
The coupling between the 2DEG and the ferromagnet is approximated by
\begin{equation}
V^{\text{fm}}_{k,\sigma} = \int dz \left [
\phi^{\text{fm}}_{k,\sigma}(z) \right ]^* V^{\text{fm}}_\sigma(z)
\chi^{\text{si}}(z)~,
\end{equation}
which is basically an overlap integral with an exponentially decaying
term.  Eq.~(\ref{Hamiltonian}) is an approximation of the
effective-mass Hamiltonian of the Si/${\text{SiO}}_2$/ferromagnet
interface in which only the most signifcant terms in the coupling have
been retained.

From the simple tight-binding Hamiltonian, Eq.~(\ref{Hamiltonian}),
the self-energy of the 2DEG electrons due to the interaction with the
ferromagnetic gate is
\begin{equation}
\Sigma_\sigma(E) = \sum_k \frac{|V^{\text{fm}}_{k,\sigma}|^2}{E
  -\epsilon^{\text{fm}}_{k,\sigma}+i\gamma^{\text{fm}}_{k,\sigma}}~.
\label{selfenergy}
\end{equation}
To account for strong scattering in the ferromagnet, a large imaginary
part has been added to the ferromagnet states
($\gamma^{\text{fm}}_{k,\sigma}$).  Calculating the complex energy
$\tilde{E}$ which satisfies $\tilde{E}- \epsilon^{\text{si}} +
i0^+-\Sigma_\sigma(\tilde{E}) = 0$, we can evaluate both the
spin-dependent 2DEG level shift, $\Delta_\sigma(\tilde{E}) =
{\text{Re}} \left [ \Sigma_\sigma(\tilde{E}) \right ]$, and the
spin-dependent 2DEG lifetime, $\tau_\sigma(\tilde{E}) = -\hbar / 2
\text{Im} \left [ \Sigma_\sigma (\tilde{E}) \right ] $, due to the
coupling with the ferromagnet.  The spin-splitting due to the level
shift, $ | \Delta_\uparrow - \Delta_\downarrow | $, is very small
compared to the Fermi energy of the 2DEG for even the thinnest
practical oxides, and will be neglected.  On the other hand, the
spin-dependent lifetime, which represents the time for an inversion
layer electron to irreversibly "fall" into the ferromagnet, is in the
picosecond range for oxide thickness' below $10~\text{\AA}$, which
approaches the intrinsic scattering time of the silicon inversion
layer, $\tau_0 \approx 1$~ps.\cite{ando} The ratio of the lifetimes
for the two spin channels $\tau_\downarrow / \tau_\uparrow$ is roughly
a factor of 2 for all oxide thicknesses; spin-dependent transport
results, which can be harnassed for spintronics purposes.  Using the
following parameters for the Si/${\text{SiO}}_2$/Fe interface, we
calculate the scattering times to be $\tau_\uparrow = 3$~ps and
$\tau_\downarrow = 6$~ps.  For silicon, the longitudinal effective
mass responsible for the confinement is $m^\star_{\text{si,l}} = 0.91
m_0$, the transverse effective mass responsible for 2DEG transport is
$m^\star_{\text{si,t}} = 0.19 m_0$, the dielectric constant is
$\epsilon_{\text{si}} = 11.7$, and the equilibrium density of the 2DEG
is assumed to be $N_0 = 10^{12}~ {\text{cm}}^{-2}$.\cite{ando} For
silicon dioxide the effective mass is $m^\star_{\text b} = 0.3 m_0$,
the dielectric constant is $\epsilon_{\text b} = 3.9$, and the barrier
height (from the Fermi level) is $U_{\text b} = 3.2$~eV.
\cite{si02_mass} For the ferromagnet, we use $U_{\text{fm}} =
2.65$~eV, exchange splitting $\Delta = 3.9$~eV, effective mass
$m^\star_{\text{fm}} = m_0$,\cite{slon} and, for simplicity, a
wavevector-independent imaginary part of the ferromagnetic energy
$\gamma^{\text{fm}}_\uparrow = 1.1$~eV and
$\gamma^{\text{fm}}_\downarrow =0.8$~eV.\cite{hong} The electric field
in the barrier is assumed to be $E^{\text{b}}_z = -10$~MeV/cm.

Having calculated the $\hat{z}$-axis properties of the
ferromagnetically-gated silicon field-effect transistor, we now
examine the 2DEG transport in the $\hat{x}$-direction.  Due to the
electrical bias on the ferromagnet, the 2DEG current is not constant
under the gate and will leak at a rate of
\begin{equation}
\frac{d }{d x} J_\sigma = -e \left [ \frac{-N_\sigma(x)}{\tau_\sigma}
\right ]~.
\label{leakage}
\end{equation}
Note that because the flow of electrons is opposite to the current,
electrons leaking out of the 2DEG into the ferromagnetic gate implies
that a positive current flows into the 2DEG; the current will increase
under the gate.  Combined with Eq.~(\ref{current}) and the appropriate
boundary conditions (the 2DEG has its equilibrium density at the
source and drain, $N_\sigma (x_{\text{s}}) = N_\sigma(x_{\text{d}}) =
N_0 / 2$), this results in a differential equation for the
spatially-dependent, spin-dependent 2DEG density.  The new scattering
channel due to the interaction with the ferromagnet must also be
included in the Drude conductivity,
\begin{equation}
g_\sigma(x) = \frac{N_\sigma(x) e^2}{m^\star_{\text{si,t}}} \left (
\frac{1}{\tau_0} + \frac{1}{\tau_\sigma} \right )^{-1}~.
\end{equation}
The diffusion constant in two dimensions calculated using the 
above conductivity is
\begin{equation}
D_\sigma(x) =
\frac{\epsilon^{\text{si}}_{\text{F},\sigma}(x)}{m^\star_{\text{si,t}}
\left ( \frac{1}{\tau_0} + \frac{1}{\tau_\sigma} \right ) \left ( 1 -
e^{-\epsilon^{\text{si}}_{F,\sigma}(x) / k_{\text B} T} \right )} ~,
\end{equation}
where the spin-dependent Fermi level is
$\epsilon^{\text{si}}_{\text{F},\sigma}(x) = 2 \pi \hbar^2 N_\sigma(x)
/ m^\star_{\text{si,t}}$.  In the following we assume the gate length
in the $\hat{x}$-direction is approximately $1000~{\text \AA}$, so
that the source is at $x_{\text{s}} = 0$ and the drain is at $
x_{\text{d}} = 1000~{\text \AA} $, and we assume that the gate extends
$1~\mu$m in the $\hat{y}$-direction.  For concreteness,  we will assume
throughout this paper that the in-plane electric field is negative,
$E_x < 0$, such that, in the absence of leakage, a negative current
would flow through the 2DEG ( electrons flow from the
source to the drain).  In addition, we assume the confinement is
homogeneous along the channel, so that $\tau_\uparrow$ and
$\tau_\downarrow$ are constant.

\begin{figure}[]
\includegraphics[width=8cm]{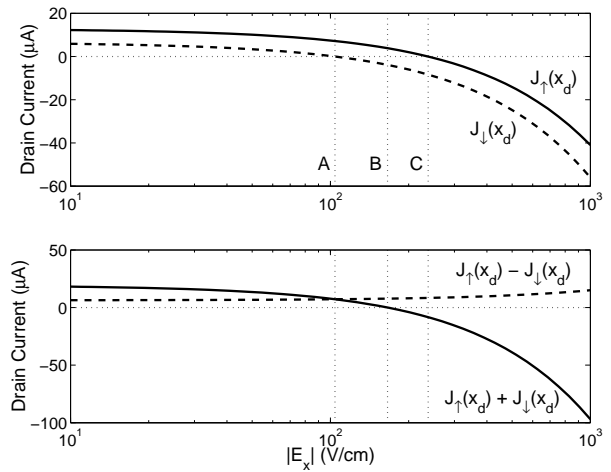}
\caption{(Top) A plot of the spin-dependent drain currents $
J_\uparrow(x_{\text d}) $ (full line) and $ J_\downarrow(x_{\text d})
$ (dashed line) as a function of the in-plane field $ E_x $.  (Bottom)
A plot of the total drain current $ J_\uparrow(x_{\text d}) +
J_\downarrow(x_{\text d}) $ (full line) and the spin current $
J_\uparrow(x_{\text d}) - J_\downarrow(x_{\text d}) $ (dashed line) as
a function of $ E_x $.  At the in-plane fields marked A and C the
drain current is $100\%$ spin-polarized; at B, a pure spin current
flows through the drain.}
\label{one_gate_field_fig}
\end{figure}

In a silicon field-effect transistor with a thick gate oxide there is
no leakage of 2DEG electrons because the wavefunction cannot penetrate
into the gate.  Neglecting the spin dependence for the time being, we
 examine the behavior of a silicon field-effect transistor with an
ultra-thin gate oxide.  Finite penetration of the 2DEG wavefunction
through the oxide and into the gate causes a tunnel current to flow.
The Fermi level in the gate is lower than the Fermi level in the
inversion layer due to the gate bias; electrons that tunnel into the
gate will inelastically fall to the gate Fermi level and have no way
to return to the 2DEG, decreasing the density in the 2DEG.  The
density will be near its equilibrium value $N_0$ at the source and
drain contacts, where electrons are injected into the 2DEG.  The
density will be lowest near the center of the gate, furthest from the
source and drain.  With no source-drain bias the density profile will
be symmetric about the center of the gate; electrons will flow into
the 2DEG from both the source and the drain, such that the drain
current is positive.\cite{momose} Application of a source drain bias
will break this symmetry, causing a drift current to flow in addition
to the diffusive current due the electron leakage.  For negative
in-plane field $E_x < 0$, at some critical field the drift current
will exactly cancel the diffusive backflow at the drain; the net drain
current will be zero.  As the in-plane field is made more negative the
drain current becomes negative, such that in the high source-drain
bias regime the diffusive current is negligible.

With a ferromagnetic gate, the above analysis still holds for each
spin channel separately (provided that the spin-flip time is much
longer than the transit time through the device).  The lifetime for
2DEG electrons to fall into the gate is spin-dependent due to the
spin-dependent coupling in the Hamiltonian, Eq.~(\ref{Hamiltonian}).
Importantly, the diffusive currents for spin $\uparrow$ electrons and
spin $\downarrow$ electrons must also be different at zero
source-drain bias, and hence the in-plane field required to exactly
cancel the diffusive backflow at the drain contact will be different
in the two spin channels.

\begin{figure}[]
\includegraphics[width=8cm]{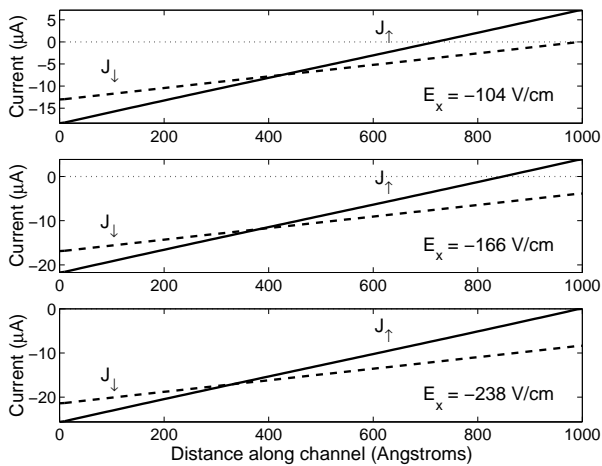}
\caption{Plots of the spin-dependent currents $J_\uparrow$ (full line)
and $J_\downarrow$ (dashed line) under the ferromagnetic gate near
field A (top), field B (middle), and field C (bottom), as labeled in
Fig.~\ref{one_gate_field_fig}.}
\label{one_gate_current_fig}
\end{figure}

The spin-dependent drain currents $J_\uparrow(x_{\text d})$ and
$J_\downarrow(x_{\text d})$ are plotted as functions of the 
in-plane field $|E_x|$ in the top panel of
Fig.~\ref{one_gate_field_fig}.  The total current $J_\uparrow(x_{\text
d})+J_\downarrow(x_{\text d})$ and the spin current
$J_\uparrow(x_{\text d})-J_\downarrow(x_{\text d})$ are plotted as
functions of $|E_x|$ in the bottom panel of
Fig.~\ref{one_gate_field_fig}.  At very low source-drain bias the
drain currents are positive in both spin channels, but have different
magnitudes due to the different leakage rates for the two spin
channels; because $\tau_\uparrow < \tau_\downarrow$, more spin
$\uparrow$ electrons leak than $\downarrow$ electrons.  The larger
gradient in the density in the spin $\uparrow$ channel results in
$J_\uparrow(x_{\text d}) > J_\downarrow(x_{\text d})$.  At the
in-plane field marked A, the drain current in the $\downarrow$ spin
channel is $J_\downarrow(x_{\text d}) = 0$; the diffusive current is
exactly cancelled by the drift current.  The larger diffusive current
in the spin $\uparrow$ channel is still of higher magnitude than the
drift current.  Spin $\uparrow$ electrons will flow into the 2DEG at
the drain contact, resulting in a $100\%$ spin-polarized current.  At
the in-plane field marked B in Fig.~\ref{one_gate_field_fig}, the
drain currents in the two spin channels are equal and opposite,
$J_\uparrow(x_{\text d}) = - J_\downarrow(x_{\text d})$.  No net
charge current flows through the drain, but a pure spin current flows
such that spin $\uparrow$ electrons are transferred into the 2DEG and
spin $\downarrow$ electrons are transferred out of the 2DEG at the
drain, i.e., the silicon inversion layer acts as a spin pump at the
field marked B.  At the field marked C the drain current in the spin
$\downarrow$ channel is negative, while in the spin $\uparrow$ channel
the drift and diffusive currents exactly cancel, $J_\uparrow(x_{\text
d}) = 0$.  Spin $\downarrow$ electrons will flow out of the the 2DEG,
resulting in a $100\%$ spin-polarized drain current.  As the magnitude
of the in-plane field is increased further the electrical currents in
both spin channels are negative; the spin $\downarrow$ channel carries
more current because its conductivity is higher than that of the spin
$\uparrow$ channel.  Electrons flow out of the 2DEG drain with some
polarization.  Note that the roles of the spin channels can be
exchanged by reversing the gate magnetization, such that the drain
contact of this ferromagnetically-gated silicon inversion layer can be
used as a source for both fully spin-polarized charge currents and
pure spin currents with either spin $\uparrow$ or $\downarrow$.  For
positive in-plane electric fields $E_x > 0$, the same phenomena happen
at the source contact.

The behavior of the spin-dependent currents throughout the 2DEG are
plotted in Fig.~\ref{one_gate_current_fig} near the in-plane fields 
A (top panel), B (middle panel), and C (bottom panel).  The leakage of
electrons under the gate is evident, as the current is not constant and 
increases under the gate.  The plots show that in this region of 
source-drain bias the drift current is just beginning to overtake the
diffusion current at the drain contact (the drift and diffusion currents
flow in the same direction at the source contact).  

The spin effects present in the silicon inversion layer with a single
ferromagnetic gate could be very attractive for spintronics
applications.  We will now describe a simple all-electrical
measurement to take advantage of the spin-dependent 2DEG transport.
Recently we proposed a planar semiconductor spin-valve which relies
upon the coupling of the inversion layer electrons to sequential
ferrromagnets.\cite{apl,prb} The space between the ferromagnets is
filled with a paramagnetic metal to ensure homogeneous 2DEG
confinement throughout the device.  A diagram of the device is shown
in Fig.~\ref{device_fig}(b).  The two ferromagnets are patterned with
different aspect ratios, such that the magnetization of one
ferromagnet can be reversed by an external field without affect the
magnetization of the other ferromagnet.

\begin{figure}[]
\includegraphics[width=8cm]{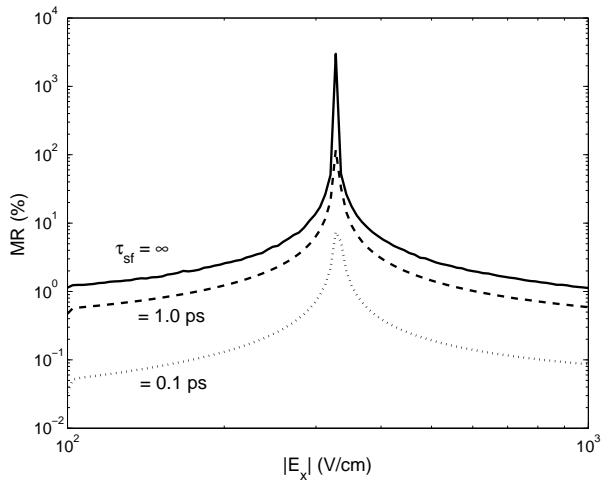}
\caption{A plot of the magnetoresistance ratio ($\%$) as a function of
the in-plane field $|E_x|$ for the spin-valve with two ferromagnetic
gates, for different values of the spin-flip time $\tau_{\text{sf}}$.  For
$\tau_{\text{sf}} = \infty$, the peak of the $MR$ is limited only by the 
number of points in the calculation.}
\label{mr_fig}
\end{figure}

Assuming the two spin channels are decoupled throughout the device,
let us examine the total 2DEG current at the drain contact for
parallel ferromagnet magnetizations, $J_{\text p}(x_{\text d})$.  One
spin channel will see the spin $\uparrow$ scattering time
$\tau_\uparrow$ under both ferromagnets, and the other spin channel
will see the spin $\downarrow$ scattering time $\tau_\downarrow$ under
both ferromagnets; the total current $J_{\text p}(x_{\text d})$ will
be the sum of the currents in the two spin channels.  Switching the
magnetization of the second ferromagnet, one spin channel will see the
spin $\uparrow$ scattering time $\tau_\uparrow$ under the first
ferromagnet and the spin $\downarrow$ scattering time
$\tau_\downarrow$ under the second ferromagnet, while the other spin
channel will see the reverse.  Because the density and current must be
continuous in both spin channels, the total drain current for
antiparallel magnetizations $J_{\text{ap}}(x_{\text d})$ will in
general be different from $J_{\text p}(x_{\text d})$.  The net result
is a {\it magnetoresistance} effect, in which the 2DEG current is
dependent upon the relative orientation of the ferromagnet
magnetizations.  We define the {\it{magnetoresistance ratio}} as
\begin{equation}
MR = 100 \cdot | \frac{J_{\text p}(x_{\text d}) -
J_{\text{ap}}(x_{\text d})} {J_{\text p}(x_{\text d}) }|~.
\label{MR}
\end{equation}
This quantity is plotted as the full line in Fig.~\ref{mr_fig} as a
function of the in-plane electric field $|E_x|$, for an
infinite spin-flip time (in which case the peak of the $MR$ is limited only
by the number of data points in the plot).  Both ferromagnets are 
assumed to be $1000~{\text{\AA}}$ long in the $\hat{x}$-direction, with
a $200~{\text{\AA}}$ gap between them.  The magnetoresistance
ratio grows to values greater than $100\%$ when $J_{\text p}(x_{\text d})
\rightarrow 0$, at which point a pure spin current would flow in the parallel
configuration.

The spin-flip time $\tau_{\text{sf}}$ couples the two spin channels
together, driving them to the same density.  Short spin flip times
($\tau_{\text{sf}}$ comparable to the scattering times $\tau_0$
and $\tau_\sigma$) will wash out any spin-dependent transport effects.
For coupled spin channels, Eq.~(\ref{leakage}) must be replaced with
\begin{equation}
\frac{d}{dx} J_\sigma = -e \left [ - \frac{N_\sigma(x)}{\tau_\sigma}
- \frac{N_\sigma(x)-N_{\bar{\sigma}}(x)}{\tau_{\text{sf}}}
\right ] ~ ,
\label{leakage_sf}
\end{equation}
where ${\bar{\sigma}}$ is spin opposite to $\sigma$.
The magnetoresistance ratio is plotted for two values of the spin-flip
time near the intrisic scattering time, $\tau_{\text{sf}} = 1.0$~ps
(dashed line) and $\tau_{\text{sf}} = 0.1$~ps (dotted line) in
Fig.~\ref{mr_fig}.  Short spin-flip times drastically reduce the $MR$;
the currents in the two spin channels are so strongly coupled that the
total current through the 2DEG is only slightly altered when switching
from parallel to antiparallel magnetizations.  Although short spin
relaxation times should not be a problem in silicon, 
\cite{portis, feher} it could be an
issue in other semiconductors.

In conclusion, we have shown how the unavoidable issue of 2DEG
carrier leakage in silicon field-effect transistors with ultra-thin gate
oxides can be put to good use for spintronics applications.  By 
replacing the metal gate with a ferromagnet, the coupling between
the silicon inversion layer and the gate becomes spin-dependent,
resulting in spin-dependent 2DEG transport.  In the low source-drain
bias regime, the drift current (first term on right side of Eq.~(\ref{current}))
and the diffusion current (second term on right side of Eq.~(\ref{current}))
are of the same order; the spin-dependent interplay between the two
results in a variety of spin currents that flow through the contacts to
such a device.  We have shown how the ferromagnetically-gated
silicon inversion layer can produce both fully spin-polarized charge
currents and pure spin currents, without any net charge transfer.  In
addition, we have shown that if the gate is made up of two sequential
ferromagnets, a magnetoresistance effect occurs in the 2DEG
current dependent upon the relative orientation of the two ferromagnet
magnetizations.  The magnetoresistance ratio can exceed $100\%$
in this low source-drain bias regime.

\end{document}